\documentclass[aps,prl,twocolumn,superscriptaddress,groupedaddress]{revtex4-1}  
\usepackage{graphicx}  
\usepackage{dcolumn}   
\usepackage{bm}        
\usepackage{amssymb, amsmath}   
\usepackage{enumerate}
\usepackage{slashed}
\usepackage{simplewick}
\usepackage{comment}
\usepackage{color}
\usepackage{soul}

\newcommand{\nn}{\nonumber}

\newcommand{\eq}[1]{(\ref{#1})}
\renewcommand{\>}{\rangle}
\newcommand{\la}{\label}
\newcommand{\ba}{\begin{align}}
\newcommand{\ee}{\end{equation}}
\newcommand{\be}{\begin{equation}}

\def\12{\frac{1}{2}}
\newcommand{\p}{\partial}
\newcommand{\en}{\end{align}}

\newcommand{\<}{\langle}

\usepackage{color}

\def\blue{\color{blue}}

\begin{document}

\setcounter{secnumdepth}{-1} 






\title{Fractional Quantum Hall Effect in a Curved Space: \\ Gravitational Anomaly and Electromagnetic Response   }
\author{T. Can, M. Laskin and P. Wiegmann}
 \affiliation{ Department of Physics, University of Chicago, 929 57th St, Chicago, IL 60637, USA}

\date{\today}
%

\date{\today}

\begin{abstract}
We develop a general method to compute  correlation functions of fractional
quantum Hall (FQH) states on a curved space. In a curved space, local transformation properties of  FQH states are examined through local  geometric variations, which are essentially governed by the gravitational anomaly. Furthermore, we show that  the  electromagnetic response  of  FQH states is related to the gravitational  response (a response
to curvature). Thus, the gravitational anomaly is also seen
in the structure factor and the Hall conductance in flat space.
The method is based on iteration of a Ward identity obtained for FQH states.
  
\end{abstract}

\date{\today}

\maketitle



\paragraph{ Introduction}

Important universal  properties of  fractional quantum Hall (FQH) states  are evident in the quantization of kinetic coefficients in terms of the filling fraction. The most well-known kinetic coefficient is the  Hall conductance \cite{Laughlin}, a transversal response to the electromagnetic field. Beside this, FQH states possess a  richer structure evident through their response to changes in spatial geometry and topology, both captured by  the gravitational response. 

A kinetic coefficient which reflects a transversal response to the gravitational field   is the odd  viscosity (also referred as anomalous viscosity, Hall viscosity or Lorentz shear modulus) \cite{Avron,Read09,TokatlyVignale}. This coefficient also 
 exhibits a quantization and reveals universal features of FQH states as much as the Hall conductance. While  the Hall conductance is  seen in  an adiabatic response to homogeneous flux deformation \cite{Laughlin}, the anomalous viscosity
 is seen as an adiabatic
response to homogeneous metric deformations
\cite{Avron}. However, even more universal features become apparent when one considers the adiabatic response to inhomogeneous  deformations of the
flux and the metric. This is the subject of this paper.

In this paper, we compute the response of the FQH states to local curvature and show that this response reveals corrections to physical quantities in a flat space that remain hidden otherwise. We compute the particle density through a gradient expansion in local curvature, explain the relation of the leading terms to the gravitational anomaly, and show that they are geometrical in nature. For this reason, we expect these terms to be universal (i.e. insensitive to the details of the underlying electronic  interaction as long  as the interaction gives rise to the FQH state). We develop a general method to compute these terms. Additionally, we show that  the dependence on  curvature determines the  long wavelength expansion of the  static structure factor in a flat background,  linking the electromagnetic response to the gravitational anomaly. Furthermore, correlation functions computed on arbitrary surfaces provide information about the properties of FQH states under general covariant and, in particular, conformal transformations.

We consider only Laughlin states for which the filling fraction \(\nu\) is the inverse of an integer. We restrict
our analysis to FQH states without boundaries.  Though our analysis is limited  to the Laughlin wave function,   we believe that our results capture the geometric properties of FQH states. As such, they may serve as universal bounds for response functions in
realistic materials exhibiting the FQH effect. Generalization of our results to other FQH is possible and will be reserved for a subsequent paper. We start by formulating the main results.
\paragraph{Main Results} 
 
We consider electrons placed on a closed oriented curved surface, such as a deformed sphere, and assume that the magnetic flux through a differential volume element of the surface \(d\Phi = B dV\) is uniformly proportional to the volume, where  \(B>0\) is a uniform magnetic field. The total number of flux quanta \(N_\phi =V (2 \pi l^2)^{-1}
\) piercing the surface is an integer equal to the area \(V\) of the surface in units of \(2\pi l^2\), where  \(l= \sqrt
{\hbar / e B}\) is the magnetic length.


In this setting, the lowest Landau level (LLL) remains degenerate on a curved surface, with a gap to excitations on the order of the cyclotron energy  \cite{Iengo}. The degeneracy of the level is determined by the Riemann-Roch theorem. Assuming
that the surface possesses no singularities so that the Euler characteristic \(\chi\) is an even
integer, the degeneracy is \(N_1=N_\phi+ \chi /2\) \cite{maraner, Iengo}. If the number of particles is exactly equal to \(N_1\), the ground state will form a droplet without a boundary, completely covering the surface.

This result readily extends to Laughlin states (for a sphere and torus see 
\cite{FDMH}, for a general Riemann surface see \cite{WenZee,FrohlichStuder,Iengo}):
the droplet has no boundary if the number of particles \(N\) is equal to
 \be N_\nu = \nu N_\phi+ \chi/ 2,\la{1}\ee  
 assuming that \(N_\nu\) is integer. We consider this case.

We focus on the particle density $\rho$ defined such that \( \rho \,dV\)  is the number of particles in the volume element \(dV\). A locally coordinate invariant  quantity, the density must be expressed locally through the (scalar) curvature \(R\). In this paper, we compute the leading terms in the gradient expansion of the density of the ground state

\begin{align} \label{rho}
\langle \rho \rangle =   \rho_0 +   \frac {1} {8 \pi} R -\frac{ b}{8 \pi} (-l^2 \Delta_{g} )R ,\ \ \ b=\frac 1 3 + \frac{\nu - 1} {4\nu},
\end{align}
where \(\rho_0 = \nu(2 \pi l^2  )^{-1}\) and \(\Delta_g\) is the Laplace-Beltrami operator. We omit higher order terms in \(l^2\).  These are controlled by the short distance physics, and are at present not accessible by our methods. 
 
The first two terms are a local version of the global relation \eq{1} between the maximum particle number and the number of flux quanta.  Eq. \eq{1} is  obtained by integrating \eq{rho} over the surface with the help of the Gauss-Bonnet theorem \(\int RdV=4\pi\chi\). Higher order terms do not contribute to this expression. 

 The second term indicates that particles accumulate in regions of positive curvature  and  repell from  regions of negative curvature. For example, it shows the excess number of  particles accumulating at the tip of a cone.  If the conical singularity is of the order \(\alpha>-1 \) such that the metric is locally \(|z|^{2\alpha}dz d\bar{z}\), the excess number of particles at the tip is \( -\alpha/2\). This term appears in equivalent form in \cite{WenZee, FrohlichStuder}.

The last term encodes the gravitational anomaly, which we explain in the body of the paper. We discuss its
implications below.

For the case of free electrons at integer filling \(\nu = 1\), Eq. \eq{rho}
was obtained in \cite{KlevtsovDouglas,Klevtsov13,Abanov13}. In equivalent form, it is  known in mathematical literature as an asymptotic expansion of the Bergman kernel \cite{Zelditch}. Defining the linear response to curvature as
\begin{align}
\eta=(\rho_0 l^2)^{-1}\frac{\delta\rho}{\delta R}\Big|_{R=0}, \la{3}
\end{align}  and passing
to  Fourier
modes,  Eq. \eq{rho} implies 
\begin{align}\la{4}
\eta(q) = \frac{1}{4\nu}(1 -  b q^2+ \mathcal O(q^4)), \qquad q=kl.
\end{align} 
In \cite{WL}, one of the authors argued that the  kinetic coefficient defined by \eq{3} enters  the
 hydrodynamics of a FQH incompressible quantum liquid (see also \cite{PW13})
as the anomalous term in the momentum flux tensor representing kinematic odd-viscosity.
The homogeneous part of the odd-viscosity,  computed through alternative methods
in \cite{Avron, Read09, ReadRezayi11, TokatlyVignale}, corresponds to the first term in \eq{4}. The leading gradient correction to the odd-viscosity for  the integer case \(\nu=1\)  was recently  computed in \cite{Abanov13}. It is related to the second term in Eq. \eq{4},  and as we show below, receives a contribution from the gravitational anomaly.

We show that the following general relation between $\eta(k)$ and the static structure factor \(s(k) = \langle \rho_k \rho_{-k} \rangle_c / \rho_0 \) is valid for Laughlin states, and likely for more general FQH states as well 
\begin{align}\label{eta}\frac {q^4} {2} \eta(q) = 
\Big( 1 + \frac {q^2 } 2 \Big ) s(q)- \frac{q^2 } 2 , \quad q=kl.\end{align}
Using these relations we obtain
\begin{align}
s(q)=   \frac 12q^{2} +  s_2q^{4} + s_{3} q^{6} + \mathcal{O}(q^{8})\la{6}
\end{align}
where 
$$s_2=(\nu^{-1}-2)/8,\quad s_3= (3\nu^{-1}- 4)(\nu^{-1} - 3)/96.$$

  The term of order \(q^4\) in the structure factor goes back to \cite{GMP}.  We find that it is controlled by $\eta(0)$ and $\lim_{q \to 0} s(q)/q^{2}$. The next correction $s_{3}$ was recently obtained in \cite{Kalinay} by means of a Mayer expansion. We provide an alternative derivation which emphasizes its connection to the gravitational anomaly.  Curiously, \(s_2\)  vanishes at \(\nu=1/2\), the bosonic Laughlin state,
while \(s_3\) vanishes for the Laughlin state at $\nu=1/3$ filling. 

We also mention another general relation between the structure factor and
the Hall conductance valid for the Laughlin wave function    
\begin{align}\la{7}
\sigma_{H}(k) & =\frac{e^{2}}{\hbar} \frac{2 \rho_{0} }{k^{2}} s(k)
\end{align} 
We clarify it in the body of the paper (see also \cite{PW13}). This
relation links the Hall conductance to the response to curvature through \eq{eta}  \cite{H}.  Furthermore, knowledge of $s_{3}$ determines the Hall conductance  up to order $k^{4}$. 

These results follow from iteration of a Ward identity obtained for the Laughlin wave function in \cite{WZ},  combined with the gravitational anomaly. An important ingredient of the Ward identity is the two point function of the ``Bose" field \(\varphi\) at merged points. The Bose field is defined as a potential of charges created by particles through the Poisson equation \begin{align}\label{poisson}-\Delta_g \varphi = 4 \pi \nu^{-1} \rho.\end{align} 
We show that in the leading $1/N$ approximation, the Bose field has Gaussian correlations. This means that (i) the connected correlation function of $\varphi$ at large distances between points is the Green function of the Laplace-Beltrami operator 
$$\Delta_g G(z,z')=- 4\pi \left[\frac{1}{\sqrt g}\delta^{(2)}(z-z')-\frac{ 1}{ V}\right],$$
 and that (ii) at small distances between points the correlation function is the regularized Green function, \begin{align} \label{G}
 \langle \varphi(1) \varphi(2) \rangle_c=\frac 1\nu \left\{
  \begin{array}{l l}
     G(1,2) &  \text{at large separation}\\
     G_R(1,2) &  \text{at short distances. }
  \end{array} \right.
\end{align}
The regularized Green function is defined as
\begin{align}\la{R}G_R(1,2)=G(1,2) + 2 \log d(1,2),\end{align} where 
\(d(1,2)\) is the geodesic distance between the  points. 

The apparent metric dependence of the two point correlation function at short distances is referred to as the gravitational anomaly.
\paragraph{ The Laughlin State on a Riemann Surface}
 It is convenient to work in holomorphic coordinates  where the  metric is conformal to the Euclidean metric \(ds^2 = \sqrt g dz d\bar z\).   In these coordinates, the scalar curvature reads \(R = -\Delta_g\log \sqrt g\), where the Laplace-Beltrami operator takes the form \(\Delta_g=(4/\sqrt g)\p\bar\p\). The K\"ahler
potential  \(K\), defined through the equation   \(\partial \bar \partial K =
\sqrt g\), also  plays an important role.

 
We choose a gauge potential with the anti-holomorphic component \(\bar{A}  =\frac 12(A_1+iA_2)=  i  B  \bar\partial K /4\), such that \(\nabla \times \mathbf A = B \sqrt g\), where $B$ is a uniform magnetic field. The states in the LLL are defined as zero modes of the antiholomorphic component of the covariant momentum 
  \begin{align} \label{pi}
  \bar\Pi =  -i \hbar \bar \partial -  e \bar A .  
  \end{align}
The  solutions to \(\bar{\Pi}\psi_{n} = 0$ are the single particle eigenstates given by \(\psi_n(z) = s_n(z) e^{-  K(z,\bar
z)/ {4 l^2}}\), where the  functions  \( s_n\) are called  holomorphic sections,  defined as solutions to  \(\bar \partial s_n = 0\) such that \(\psi_n\) is normalizable.

The many-body  ground state wave function for free fermions is the Slater determinant of the single particle eigenstates \(\Psi_1 ( z_1, ..., z_N) \propto e^{- \sum_i^N K(z_i,\bar z_i) /4l^2} \det [s_n(z_i) ]\), as shown in \cite{Iengo, Klevtsov13}.


For simplicity, we specialize to a surface of genus zero, with a marked point chosen to be at infinity where $K \sim 
(V /\pi) \log |z|^2+{o}(1 )$  and $\log\sqrt g\sim -2\log |z|^2$. In this case, the holomorphic sections
 \(s_n(z)\) are polynomials of degree  \(n = 0, 1, ..., N_{\phi}\), and the Vandermonde identity implies \(\det
[s_n(z_i) ]\propto\prod_{i < j}^N
(z_i - z_j) \).

We construct Laughlin states at the filling fraction \(\nu \)  by raising the determinant to the power equal to the inverse filling fraction $\beta \equiv \nu^{-1}$. On a surface of genus zero, the $N$ electron Laughlin wave function is
\begin{align}\label{wf}
\Psi_{\beta} = \frac 1 {\sqrt {{\mathcal Z}[g]}} \prod_{i < j}^N (z_i - z_j)^\beta e^{-\frac{1}{4 l^2} \sum_i^N K(z_i,\bar z_i) }
\end{align}
where \({\mathcal Z}[g]\) is a normalization factor. This wave function is normalizable
 only for \(N \leq N_\nu\) given by \eq{1}. We consider states with \(N=N_\nu\), the only case in which the wave-function is modular invariant, indicating the electron droplet completely covers the surface. 

Though we work only in genus zero, our formulas are local and therefore apply to more general surfaces. 
For a comprehensive discussion of the LLL on a surface of arbitrary genus see \cite{Iengo}.

 In the case of a sphere of radius $r$, inserting the K\"ahler potential \(K= 4r^2 \log (1 + |z|^2 / 4r^2 ) \) into \eq{wf} reproduces the well-known  wave-function on a sphere in stereographic coordinates \cite{FDMH}. In the limit that \(r \rightarrow \infty\), \(K = |z|^2\) and the planar wave function is recovered.

 With this setup, we wish to evaluate equal time correlation functions in the limit \(l \rightarrow 0, N_\phi \rightarrow \infty\)  such that the area \( V = 2 \pi l^2 N_\phi \) is fixed.
\paragraph{Generating functional  }
 The normalization factor \(\mathcal Z[ g]$ encodes the geometry of the surface through its dependence on the metric. It can be used to generate response functions to  deformations of the metric. From (\ref{wf}) the generating functional is defined as 
\begin{align}\label{Z}
\mathcal Z[g] = \int \prod_{i < j}^N |z_i - z_j|^{2\beta} \prod_i^N e^{  W(z_i,\bar
z_i)} d^2 z_i, 
\end{align}
where $W = - K/2l^{2} +  \log \sqrt{g}$. Each variation of $\log\mathcal{Z}$ over $W(z, \bar{z})$ inserts a factor of $ \sqrt{g(z)} \rho(z)$ into the integral \eq{Z}, where 
$$ \rho (z)  = \frac{1}{\sqrt g} \sum_i \delta^{(2)} (z-z_i),$$
is the particle density. Higher order connected correlation functions of the density can be generated in this manner. 

More generally, if  \(\mathcal{A}(z_1,...,z_N)\) is a (metric independent) symmetric function of the coordinates, then
 \begin{align}\la{rhoW} 
 \delta \langle\mathcal{A}\rangle/\delta W=\sqrt g\langle\mathcal{A}\,\rho \rangle_c .
 \end{align}
 This method for computing correlation functions is detailed in \cite{WZ}.
\paragraph{ Relations between linear responses on
the lowest Landau level}
Using the explicit dependence of \(W\) on \(\sqrt g\), we observe a general relation for a linear response to area preserving  variations of the metric 
\begin{align}\la{171}\frac 12(-l^2\Delta_g)\frac{\delta\langle \mathcal{A}\rangle}{\delta \sqrt{ g(\zeta)}} =\Big ( 1 + \frac 12(- {l^2} \Delta_g)\Big)\langle  \mathcal{A} \rho(\zeta)\rangle_c. \end{align} 
This relation is valid for any \(N\) and any \(\beta\)  (including  the integer case). With the choice of \(\mathcal{A}= \sum_i \delta^{(2)}(z-z_i)\) and the identity \(\delta \langle \rho\rangle/ \delta \sqrt{g}|_{R = 0} = - \Delta \left[ \delta \langle \rho \rangle / \delta R \right|_{R = 0}$, we obtain \eq{eta}.

This relation reflects a symmetry between gravity and electromagnetism specific
to the LLL. It can be traced back to properties of zero
modes  of the operator (\ref{pi}). 

Similar arguments lead to the relation between the static structure factor and
the Hall conductance  expressed in \eq{7}. The generating functional  \eq{Z}
can be seen as the normalization factor  of the  Laughlin wave function in a flat space, but in a weakly inhomogeneous magnetic field. A key assumption is that  the form of the wave function is the same  as  in the case of a uniform magnetic field where \(B=-(\hbar/2e)\Delta W\),  as in \cite{ABWZ}. The two-point density correlation function 
\begin{align}\nn
\langle \rho(z)\rho(z')\rangle_c = \delta\<\rho(z)\>/\delta W(z'),
\end{align}
can then be connected to a variation of the density over magnetic field at fixed filling fraction. In Fourier modes, this identity becomes 
\begin{align}\nn
\rho_0s(k)=(\hbar/2e)k^2(\delta\rho_k/\delta B_k).
\end{align}
The  inhomogeneous version of the Streda formula
\(e\delta\rho_k/\delta B_k=\sigma_{H}(k) \),
 yields the relation \eq{7}  \cite{comment}. Then, computing \(\eta(k)\) allows us to extract \(s(k)\), and thus \(\sigma_{H}(k),
\) from (\ref{eta}). Moreover, once we compute \(\langle\rho\rangle\), we can  recover the generating functional which we present in the end of the paper.
 
To compute the response to  curvature we employ the Ward identity   explained in the next section. 
\paragraph{ Ward identity}
The generating functional \({\mathcal Z}[g]$ is invariant under any transformation
of coordinates of the integrand \eq{Z}. In particular, a holomorphic infinitesimal
diffeomorphism $z_i \rightarrow z_i + \epsilon / (z - z_i) $ where $z$ is a parameter, invokes
a change of the integrand \eq{Z} by the factor 
\begin{align}
\sum_i\frac{\p_{z_i} W}{z-z_i}+\sum_{j\neq i}\frac{\beta}{(z-z_i)(z_i-z_j)}+\sum_i\frac{1}{(z-z_i)^2}.\nonumber
\end{align}
The Ward identity
states that the expectation  value of this factor vanishes. Expressing the sum as an integral over the density \(\sum_i\to\int d^2\xi\sqrt {g(\xi)}\rho(\xi)\), yields the
relation connecting one- and two-point  functions 
 \begin{align}\label{loopint}
 -2  \beta \int \frac{  \partial W}{z - \xi} \langle \rho  \rangle  \sqrt g d^2 \xi  = \langle (\partial \varphi )^2 \rangle + (2 - \beta)  \langle \partial^2 \varphi  \rangle, 
\end{align}
where the Bose field \(\varphi=-\beta\sum_i\log|z-z_i|^2\).
 Eq. \eq{loopint} was obtained in \cite{WZ}. Furthermore, it is convenient to define the field $$\tilde \varphi  = \varphi + \frac K {2 l^2} - \frac \beta 2 \log \sqrt g,$$
that vanishes at \(z\to \infty\). The anti-holomorphic derivative of Eq.(\ref{loopint}) eliminates the integral, by virtue of the \(\p\)-bar formula $\bar{\partial(} \frac{1}{z} )= \pi \delta^{(2)}(z)$, to give
\begin{align} \label{loopdiff}
\langle  \rho \rangle\partial \langle  \tilde \varphi \rangle  +   \Big ( 1 - \frac{\beta}{2} \Big) \partial \langle \rho \rangle = \frac{1}{2\pi \beta  \sqrt{g}}  \bar \partial \langle ( \partial \tilde\varphi)^2 \rangle_{c}.
\end{align}
\paragraph{Iterating the Ward identity: the leading order} 
The Ward identity consists of terms of different orders in \(N\), and 
can be solved iteratively order by order. The first term on the l.h.s. of \eq{loopdiff} is of the
order \(N^2\), the other two  are of the order \(N\). To leading order we thus have $\langle \tilde \varphi \rangle= 0 $, which yields
\begin{align}
\langle \varphi \rangle =- \frac K {2 l^2} + \frac \beta 2 \log \sqrt
g+ \mathcal O (l^2).
\end{align}
From this, using \eq{poisson} we recover the first two terms in  \eq{rho}.

To proceed with the next iteration, we need to know \(\langle ( \partial \tilde\varphi)^2 \rangle_{c}  $
or rather the short distance behavior of the connected two-point correlation function \(\langle \varphi(z) \varphi(z') \rangle_c$.  

\paragraph{The Gravitational Anomaly}
We obtain the two-point function from \eq{rhoW} by varying the one-point function of \(\varphi\) with respect to \(W\), which to leading order implies 
\begin{align}\nn
\Delta_{g}\langle \varphi (z)\varphi(z')\rangle_c
=  -4\pi \beta\left[ \frac{1}{\sqrt g}\delta^{(2)}(z\!-\!z')-\frac 1V\right],
\end{align}
where we made use of \eq{poisson} to rewrite $\rho$ in terms of $\varphi$. Thus, we see that the two-point function is the Green function of the
Laplace-Beltrami operator as in \eq{G}.
At short distances, the two-point  correlation function \(\langle
\varphi (z)\varphi(z')\rangle_c\) is regular, and general covariance requires regularization to be as in Eq. (\ref{G}). We save further discussion of this subtle point for a subsequent paper. 

We are now in a position to compute the missing ingredient \ of the Ward
identity (\ref{loopdiff}). Taking  derivatives and merging points,
we obtain the known result for the Green function
\begin{align}\nn
\langle (\partial \tilde\varphi(z))^{2}\rangle_{c} 
 = \frac{\beta}{6} \left[\partial^{2} \log \sqrt{g} - \frac{1}{2}\left( \partial \log \sqrt{g}\right)^{2} \right].
\end{align}
Applying $\bar{\partial}$ we obtain the {\it anomalous} part of the Ward identity \eq{loopdiff},
 \begin{align}\label{rhsloop} \frac 1 {\sqrt g} \bar   \partial \langle (\partial \tilde\varphi(z))^{2}\rangle_{c}= -\frac{\beta}{24} \partial R,    \end{align}
This formula represents the {\it gravitational} or {\it trace anomaly}. It shows that the connected correlation function of the holomorphic derivative of the Gaussian field is no longer holomorphic in a curved space.
\paragraph{ Iterating the Ward identity: subsequent orders}
The anomalous contribution (\ref{rhsloop}) allows us to extract $b$ by computing the next order in the Ward identity. Inserting (\ref{rhsloop}) into (\ref{loopdiff}), matching terms of the same order (by replacing the first  \(\<\rho\>\) in (\ref{loopdiff}) with its leading order) reduces \eq{loopdiff} to linear form, which readily integrates to
\begin{align}\label{loop2} \rho_0\langle  \tilde \varphi \rangle  +   \Big (
1 - \frac{\beta}{2} \Big)  (\langle \rho\rangle-  \rho_0) = -\frac{1}{48\pi} R.
  \end{align} 
In this equation, all of the terms are proportional to the curvature. The r.h.s. is proportional to the trace anomaly of the free Gaussian field. Matching the coefficients determines the coefficient  \(b\) in  (\ref{rho}).



  \paragraph{ Generating functional and Polyakov's Liouville action}
Once we know the density \eq{rho}, the generating functional  can be computed  by integrating \eq{rhoW} with $\mathcal{A} = \mathcal{Z}[g]$, in a similar manner to what was done in \cite{WZ}
 \begin{align}  \frac 12(-l^2\Delta_g)\frac {\delta \log{\mathcal Z}[g]} {\delta\sqrt g}&=\Big ( 1 +\frac 12 (-{l^2} \Delta_g)\Big)\<\rho\>.
\end{align} 
The result for $\beta=1$ was presented in the recent paper \cite{Klevtsov13}.  The generating functional for an arbitrary filling fraction, developed as an expansion in $1/N_\phi$, reads
 \begin{align}
\nn &\log\frac {{\mathcal Z}[g]}{{\mathcal Z}[g_0]}=\frac{N_{\phi}}{2}(N_\nu\!+\!1)+N_\phi ^2A^{(2)}[g]+N_\phi A^{(1)}[g]+A^{(0)}[g],\\ \nn
&A^{(2)}=-\frac{\pi}{2 \beta }\frac{1}{V^2}  \int K dV, \quad  A^{(1)}=\frac{1}{2V  }\int  \log\sqrt g \,dV,\\&\nn A^{(0)}= \frac{1}{16\pi
 }\left(\frac{1}{3} + \frac{\beta-1}{2}\right)\left(\int
 \log\sqrt g  R\,dV+16\pi\right),\quad   
\end{align}
where  \({\mathcal Z}[g_0]\) is the generating functional  on a sphere. 

  The functionals \(A^{(2)}\) and \(A^{(1)}\) are known objects in  K\"ahler
geometry \cite{FKZ12,Klevtsov13}. Unlike the higher order terms, the first three terms cannot be expressed locally through the scalar curvature $R$. For this reason, they obey non-trivial co-cycle properties  explained in  \cite{FKZ12,Klevtsov13}.   The variations of the first two functionals  over the K\"ahler potential 
are the volume form and
the curvature. 

The generating functional encodes the gravitational and electromagnetic response
of the FQH states. It shows how various correlation functions  transform under variations of the geometry such as conformal transformations.  In particular, the order zero (dimensionless) term $A^{(0)}$ reflects the gravitational anomaly. The functional \(A^{(0)}\)  is Polyakov's Liouville action. Recall that Polyakov's action represents the partition function of a  Gaussian free field \cite{Polyakov}. It
 appears as a normalized spectral determinant of the Laplace-Beltrami operator
\begin{align}
-\frac 12\log \frac{{\rm det }\left(-\Delta_g\right)}{{\rm det}\left( -\Delta_{g_0}\right)}=\frac{1}{96\pi}\int
 \log\sqrt g  R\,dV+\frac 16.
\end{align}
Polyakov's action transforms covariantly under conformal transformations.


\noindent\ We are grateful to A.G. Abanov, A. Cappelli, A. Gromov, I. Gruzberg, S. Klevtsov,  D. T. Son, A. Zabrodin and S. Zelditch for inputs at different stages of this work. The work  was supported by NSF DMS-1206648, DMS-1156656, DMR-MRSEC 0820054 and  John Templeton Foundation.


\begin{thebibliography}{99}
\bibitem{comment}
In realistic materials the momentum dependence of the  Hall conductance may deviate from that given by the structure factor \cite{Abanov13, SonNC}. It may include effects of diamagnetic currents caused by an inhomogeneous electronic density. 
Eq. \eq{7} captures only the universal part of the Hall conductance and may serve as a bound.
\end{thebibliography}
\end{document}